\documentclass{aa}
\usepackage{graphicx}
\usepackage{longtable}
\usepackage{ulem}

      \def\new#1 {{\bf #1 }}
      \def\cut#1 {\sout{#1} }

\begin{document}
\def\ffam {\hbox{$\,.\!\!^{\prime}$}}
\def\ffas {\hbox{$\,.\!\!^{\prime\prime}$}}
\def\ffM {\hbox{$\,.\!\!^{\rm M}$}}
\def\ffm {\hbox{$\,.\!\!^{\rm m}$}}
\def\ffs {\hbox{$\,.\!\!^{\rm s}$}}

\title{Redshifted formaldehyde from the gravitational lens B0218+357}


\author{
         N. Jethava\inst{1}
        \and
         C. Henkel\inst{1}
        \and
         K.M.~Menten\inst{1}
        \and
         C.L.~Carilli\inst{2}
        \and
         M.J.~Reid\inst{3}}

\offprints{N. Jethava, \email{njethava$@$mpifr-bonn.mpg.de}}

\institute{Max-Planck-Institut f{\"u}r Radioastronomie, Auf dem H{\"u}gel 69, D-53121 Bonn, Germany
           \and
           National Radio Astronomy Observatory, Socorro, New Mexico, 87801, USA
           \and
           Harvard-Smithsonian Center for Astrophysics, 60 Garden St., MS 42, Cambridge, MA\,02138, USA}

\date{Received date / Accepted date}

\abstract
{Little is known about individual molecular clouds at high redshifts. The gravitational lens toward B0218+357 offers the unique possibility to study cool 
moderately dense gas with high sensitivity and angular resolution in a cloud that existed half a Hubble time ago.}
{This non-CO molecular multi-level study of a significantly redshifted cloud aims at a better definition of the physical properties of molecular 
gas in a kind of interstellar environment that is rarely studied in the Galaxy or in the nearby extragalactic space.}
{Observations of the radio continuum and six formaldehyde (H$_2$CO) lines were carried out with the VLA, the Plateau de Bure interferometer
and the Effelsberg 100-m telescope.}
{Three radio continuum maps indicate a flux density ratio between the two main images, A and B, of $\sim$3.4$\pm$0.2. Within the errors the ratio is 
the same at 8.6, 14.1, and 43\,GHz. The 1$_{01}-0_{00}$ line of para-H$_2$CO is shown to absorb the continuum of image A. Large Velocity Gradient 
radiative transfer calculations are performed to reproduce the optical depths of the observed two cm-wave "K-doublet" and four mm-wave rotational lines. 
These calculations also account for a likely frequency-dependent continuum cloud coverage. Confirming the diffuse nature of the cloud, an $n$(H$_2$) 
density of $<$1000\,cm$^{-3}$ is derived, with the best fit suggesting $n$(H$_2$) $\sim$ 200\,cm$^{-3}$. The H$_2$CO column density of the main 
velocity component is $\sim$5$\times$10$^{13}$\,cm$^{-2}$, to which about 7.5$\times$10$^{12}$\,cm$^{-2}$ has to be added to also account for a weaker 
feature on the blue side, 13\,km\,s$^{-1}$ apart. $N$(H$_2$CO)/$N$(NH$_3$) $\sim$ 0.6, which is four times less than the average ratio obtained from 
a small number of local diffuse (galactic) clouds seen in absorption. The ortho-to-para H$_2$CO abundance ratio is 2.0--3.0, which is consistent with 
the kinetic temperature of the molecular gas associated with the lens of B0218+357. With the gas kinetic temperature and density known, it is found
that optically thin transitions of CS, HCN, HNC, HCO$^+$, and N$_2$H$^+$ (but not CO) will provide excellent probes of the cosmic microwave background 
at redshift $z$=0.68.}
{}

\titlerunning{Redshifted H$_2$CO toward the gravitational lens B0218+357}

\authorrunning{Jethava et al.}

\maketitle


\section{Introduction}

B0218+357 is one of the most interesting gravitational lens systems studied to date. The background source showing a red 
featureless optical spectrum (Stickel \& K{\"u}hr \cite{stickel93}) is believed to be a BL Lac object. The prominent radio source has a
redshift of $z$$\sim$0.944$\pm$0.002 (Cohen et al. \cite{cohen03}) and is lensed by an intervening almost face-on spiral (for an image, 
see York et al. \cite{york05}) into two compact images, A and B, and a faint Einstein ring (e.g., O'Dea et al. \cite{odea92}; Patnaik et al. 
\cite{patnaik93}). The flux density of source A is about three times higher than that of source B; their separation as well as the angular 
diameter of the ring are with $\sim$0\ffas 3 exceptionally small.

Browne et al. (\cite{browne93}) reported optical detections of narrow atomic absorption and emission lines from the lens at redshift 
$z$=0.6847. H\,{\sc i} absorption was observed by Carilli et al. (\cite{carilli93}) and Kanekar et al. (\cite{kanekar03}). The first 
molecular absorption lines from the intervening system were detected by Wiklind \& Combes (\cite{wiklind95}). Identified molecular 
species are CO, HCN, and HCO$^+$ (Wiklind \& Combes \cite{wiklind95}), H$_2$CO (Menten \& Reid \cite{menten96}), H$_2$O (Combes \& 
Wiklind \cite{combes97a}), CS (Combes et al. \cite{combes97b}), OH (Kanekar et al. \cite{kanekar03}), NH$_3$ (Henkel et al. \cite{henkel05}) 
and tentatively also LiH (Combes \& Wiklind \cite{combes98}). These absorption lines originate from a diffuse molecular cloud of kinetic 
temperature $T_{\rm kin}$$\sim$55\,K (Henkel et al. \cite{henkel05}) that is located along the line-of-sight to component A (Menten \& Reid 
\cite{menten96}). The cloud parameters appear to be quite peculiar when compared with the properties of clouds in the Galaxy or in nearby 
extragalactic systems (Henkel et al. \cite{henkel05}).

The gravitational lens of the B0218+357 system provides a unique laboratory for detailed molecular cloud research at intermediate redshifts. 
From the standard model ($\Lambda$-cosmology with $H_0$=71\,km\,s$^{-1}$\,Mpc$^{-1}$, $\Omega_{\rm m}$=0.27 and $\Omega_{\Lambda}$=0.73;
Spergel et al. 2003) we obtain a light travel time of 6.2\,Gyr, a luminosity distance of 4.1\,Gpc, a comoving radial distance of 2.5\,Gpc
and an angular size distance of 1.5\,Gpc, resulting in a linear scale of 7.1\,pc\,mas$^{-1}$.

In the following we present three radio continuum maps of B0218+357 as well as the first non-CO molecular multilevel study of a significantly 
redshifted cloud. The species of choice is formaldehyde (H$_2$CO). With the detection of five H$_2$CO transitions complemented by an additional 
H$_2$CO spectrum from Menten \& Reid (\cite{menten96}), we have a total of six lines (see also Darling \& Wiklind \cite{darling05}). These and 
the continuum data can be used to further elucidate the highly peculiar physical cloud parameters (Henkel et al. \cite{henkel05}), to constrain 
the continuum morphology and to estimate an ortho- to para-H$_2$CO abundance ratio that has the potential to provide clues to the history and 
chemistry of the cloud.

\begin{table*}
\caption{Summary of spectroscopic observations$^{a)}$}
\label{obs}
\begin{center}
\begin{tabular}{ccrrcccccc}
\hline
\hline
Molecule   & Line                 &\multicolumn{2}{c}{Frequency}&Telescope & Date       & Channel      & FWHM             &Position & rms          \\
           &                      &  Rest     &  Redshifted     &          &  of        & width        & beam             & angle   &              \\
           &                      &           &                 &          & observation&              & size             &         &              \\
           &                      &\multicolumn{2}{c}{(GHz)}    &          &            & (km s$^{-1})$&(arcsec)          &(degrees)&(mJy)         \\
\hline  
H$_2$CO(O) & 1$_{10}$$-$1$_{11}$  &   4.830   &   2.867         &Effelsberg& 05-Sep-97  &    1.6       & 250              &  --     &  1           \\
H$_2$CO(O) & 2$_{11}$$-$2$_{12}$  &  14.488   &   8.600         &VLA-A     & 24-Aug-95  &    1.7       & 0.26$\times$0.24 &  27     &  2           \\
H$_2$CO(P) & 1$_{01}$$-$0$_{00}$  &  72.838   &  43.236         &VLA-A     & 02-Dec-96  &    3.1       & 0.05$\times$0.04 &  16     & 10           \\
H$_2$CO(O) & 2$_{12}$$-$1$_{11}$  & 140.840   &  83.601         &PdBI-5C2  & 09-Nov-96  &    0.25      & 6.03$\times$5.29 &  11     & 36           \\
H$_2$CO(P) & 2$_{02}$$-$1$_{01}$  & 145.603   &  86.429         &PdBI-5C2  & 23-Mar-97  &    0.25      & 5.83$\times$5.06 &  15     & 31           \\
H$_2$CO(O) & 2$_{11}$$-$1$_{10}$  & 150.498   &  89.335         &PdBI-5C2  & 05-Aug-97  &    0.25      & 5.64$\times$4.39 &  14     & 35           \\

\hline   
 
\end{tabular}
\end{center}
a) Col.\,1: H$_2$CO species (O: ortho; P: para). Col.\,2: Quantum numbers of the respective transition ($J_{\rm KaKc}$). Cols.\,3 and 4: Rest and 
redshifted frequencies. Cols.\,5--7: Observatory (VLA: Very Large Array; PdBI: Plateau de Bure Interferometer), epoch and channel width. Cols.\,8 
and 9: Full Width to Half Power (FWHP) beam size and major axis beam orientation. Col.\,10: 1$\sigma$ noise level.
\end{table*}

\section {Observations}

Table~\ref{obs} and Fig.\,\ref{7353fig1} summarize the observational parameters of the H$_2$CO line measurements, including the 2$_{11}$$-$2$_{12}$ 
line reported by Menten \& Reid ({\cite{menten96}). Observations were carried out with the Very Large Array (VLA) of the National Radio Astronomy 
Observatory (NRAO\footnote{The NRAO is a facility of the National Science Foundation operated under cooperative agreement with Associated 
Universities, Inc.}), the Plateau de Bure Interferometer (PdBI) of the Institute de Radioastronomie Millim{\'e}trique (IRAM\footnote{IRAM is 
supported by INSU/CNRS (France), the MPG (Germany) and the IGN (Spain).}) and the 100-m Effelsberg telescope of the Max-Planck-Institut f{\"u}r 
Radioastronomie (MPIfR\footnote{Based on observations with the 100-m telescope of the MPIfR (Max-Planck-Institut f{\"u}r Radioastronomie) at Effelsberg, Germany.}).

\subsection{VLA}

The VLA observations were carried out in the A-configuration at 8.6 and 14.1\,GHz (X- and U-band; see Menten \& Reid \cite{menten96}) and at 
43.2\,GHz (Q-band; projects AM0494 and AM0545), employing 25, 25, and 13 25-m antennas, respectively. Measurements at the two lower frequencies, 
already presented by Menten \& Reid (\cite{menten96}), were re-reduced (note the slight difference in angular resolution of the X-band data between 
those given in our Table~\ref{obs} and that of Menten \& Reid \cite{menten96}). In the X- and U-band, the total bandwidth of 6.25\,MHz was split 
into 64 channels. In the Q-band, the bandwidth was 12.5\,MHz, consisting of 32 channels. The phase center was located at $\alpha_{\rm J2000}$ = 
02$^{\rm h}$21$^{\rm m}$05\ffs 4733 and $\delta_{\rm J2000}$ = +35$^{\circ}$56$^{\prime}$13\ffas 791. Phase and bandpass calibration were 
achieved by observations toward the standard VLA calibrator 3C\,84. Absolute flux calibration was obtained by measuring 3C\,286, for which 
flux densities of 5.11, 3.54 and 1.46\,Jy were adopted at 8.6, 14.1 and 43.2\,GHz, respectively (Ott et al. \cite{ott94}). 

The data were calibrated and reduced according to the standard NRAO Astronomical Image Processing system (AIPS). Using the AIPS task UVLIN, we 
removed the contribution of the continuum in the uv-plane to obtain the spectral line data that were imaged and CLEANed. The B0218+357 
continuum source is strong (see Sect.\,3.1) so that several iterations of self calibration could be performed to improve the image quality and 
to determine the phase and amplitude corrections which were applied to each spectral line channel. The data were converted into CLASS format 
to perform the final analysis.

\begin{figure}[t]
\vspace{-0.1cm} \centering
\includegraphics[scale=0.90,angle=00]{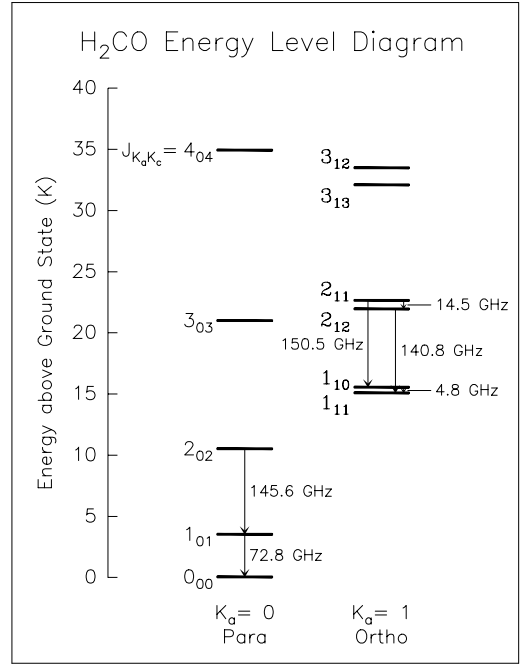}
\vspace{0.2cm}
\caption{H$_2$CO energy level diagram up to 40\,K above the ground state. Note the "K-doublets" in the $K_{\rm a}$=1 ladder, that are the result 
of the slight asymmetry of the molecule. For the six observed transitions that are marked by arrows (two $J$=2--1 ortho-H$_2$CO rotational transitions, 
one $J$=1--0 and one $J$=2--1 para-H$_2$CO rotational transition and the $J$=1 and 2 K-doublet ortho-H$_2$CO transitions), rest frequencies are given. 
\label{7353fig1}}
\end{figure}

\subsection{PdBI}

The PdBI observations were carried out in the 5C2-array configuration (project G048) at frequencies of 83.601, 86.429 and 89.335\,GHz (see 
Table~\ref{obs}). The array was equipped with five 15-m antennas. At 83\,GHz, amplitude calibration was obtained using the standard PdBI 
calibrators 0415+379 and 0234+285 and adopting flux densities of 6.28 and 1.36\,Jy. At 86\,GHz, 3C\,454.4, 3C\,84 and 0415+379 with fluxes 
of 6.14, 5.38, and 4.1\,Jy were used. Amplitude calibration at 89\,GHz was obtained measuring 2230+114, 3C\,84 and adopting fluxes of 
7.97 and 4.64\,Jy, respectively. For the phases, the standard PdBI calibrator 3C\,84 was observed. The data were reduced using "Continuum 
and Line Interferometer Calibration" (CLIC) standard procedures. The calibration tables were fed into MAPPING to subtract the continuum 
levels and to convert the data into CLASS format. Since B0218+357 is a strong continuum source (Sects.\,1 and 3.1), we could perform self-calibration, 
which significantly reduced formal phase and amplitude uncertainties.

\subsection{Effelsberg}

Observations with the 100-m telescope at Effelsberg were carried out in September 1997 using a dual channel 9\,cm wavelength HEMT receiver. 
These measurements were done in a position switching mode with a two level autocorrelation spectrometer containing 2$\times$512 channels with a 
bandwidth of 6.25\,MHz. Pointing corrections could be determined toward B0218+357 itself and were found to be smaller than 10$^{\prime\prime}$. 
Calibration was obtained from continuum cross-scans toward 3C\,286.

\begin{figure}[t]
\vspace{-0.0cm} \centering
\includegraphics[scale=0.77, angle=0]{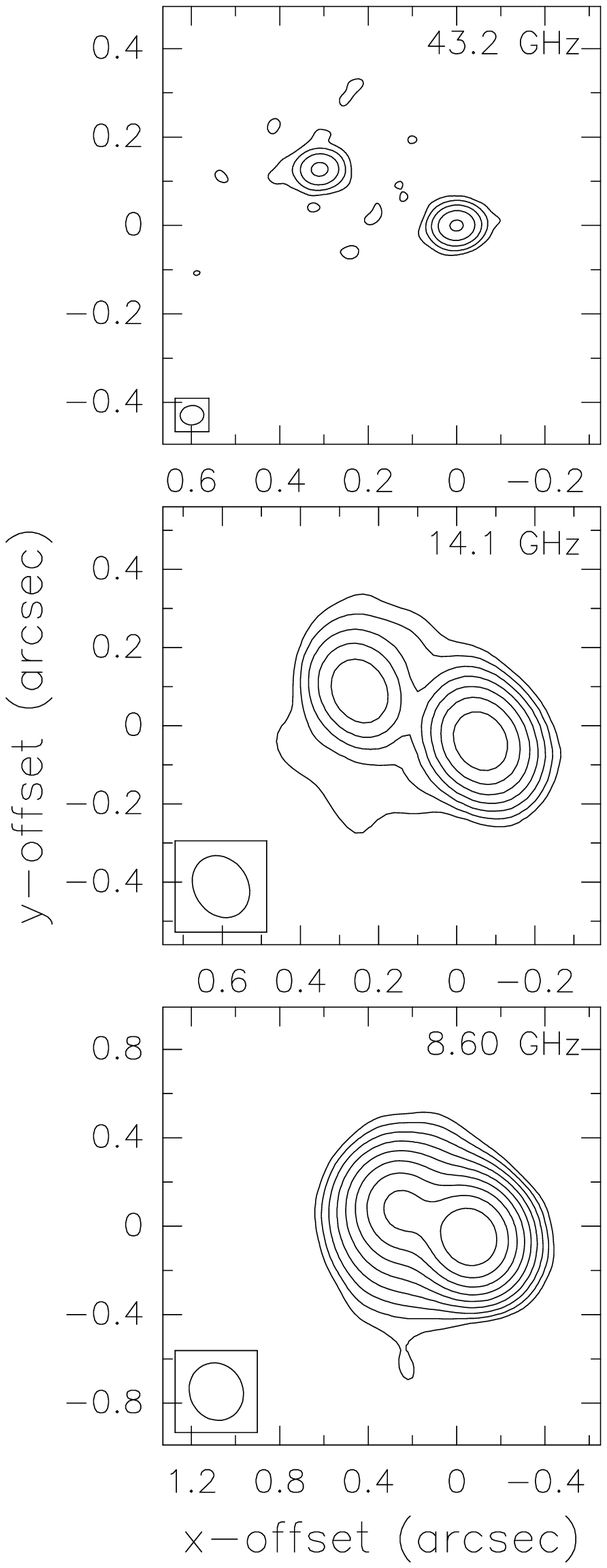}
\vspace{-0.2cm}
\caption{VLA A-array continuum images of B0218+357 taken at Q-, U-, and X-band (top to bottom). In each image, the lowest contour line represents 
four times the rms noise level, which is $\sigma$ = 0.21, 2.1, and 0.5 mJy~beam$^{-1}$ for the Q-, U, and X-band images, respectively. Contour levels 
increase by a factor of 4 for the Q-band image and by a factor of 2 for the U- and the X-band images. Peak flux densities are 0.267, 0.99, 
and 1.0 Jy~beam$^{-1}$. The full width to half maximum dimensions of the restoring beam are shown in the lower left corner of each panel. 
The origin of each image is the phase center position given in Sect.\,2.1.
\label{7353fig2}}
\end{figure}

\begin{table}
\caption{Continuum flux densities$^{a)}$}
\label{cont}
\begin{center}
\begin{tabular}{cccc}
\hline
Frequency  &  Epoch  &   Total flux    &    A/B     \\
           &         &    density  &            \\
 (GHz)     &         &     (Jy)        &            \\          
\hline   \\

  2.9      &  05-Sep-97   & 0.69$\pm$0.06   & 	     --       \\
  8.6      &  24-Aug-95   & 1.30$\pm$0.13   & 	3.3$\pm$0.13  \\
 14.1      &  24-Aug-95   & 1.18$\pm$0.12   & 	3.3$\pm$0.14  \\
 43.2      &  02-Dec-96   & 0.34$\pm$0.03   & 	3.5$\pm$0.22  \\
 83.6      &  09-Nov-96   & 0.58$\pm$0.09   & 	     --       \\
 86.4      &  23-Mar-97   & 0.48$\pm$0.07   & 	     --       \\
 89.3      &  05-Aug-97   & 0.52$\pm$0.08   &             --       \\

\hline   
 
\end{tabular}
\end{center}
a) Col.\,1: Observed frequencies; to compare with the rest frequencies given in Tables~\ref{obs} and \ref{lines}, multiply by 1.68. For details 
of the 8.6 and 14.1\,GHz masurements, see also Menten \& Reid (\cite{menten96}). Col.\,4: Flux density ratios between the two main lensed images 
of the background source.
\end{table}

\section{Results}

\subsection{Continuum}

Menten \& Reid (\cite{menten96}) presented VLA A-array continuum maps of B0218+357 at 8.6 and 14.1\,GHz. We have re-reduced their data along with 
new A-array observations at 43.2\,GHz. Fig.\,\ref{7353fig2} presents the three continuum images. The upper panel displays the 43.2\,GHz map with the two 
main radio continuum components A and B being well separated. The two components and traces of the ring (e.g. Mittal et al. \cite{mittal06}) are 
seen in the central panel. These main features are less well separated in the lower panel, where the angular resolution is not quite as good. 
Total flux densities including the Effelsberg and PdBI data as well as the VLA A-array flux density ratios between images A and B are given in 
Table~\ref{cont}.

\begin{figure}[t]
\vspace{-0.0cm} \centering
\includegraphics[scale=.92,angle=0]{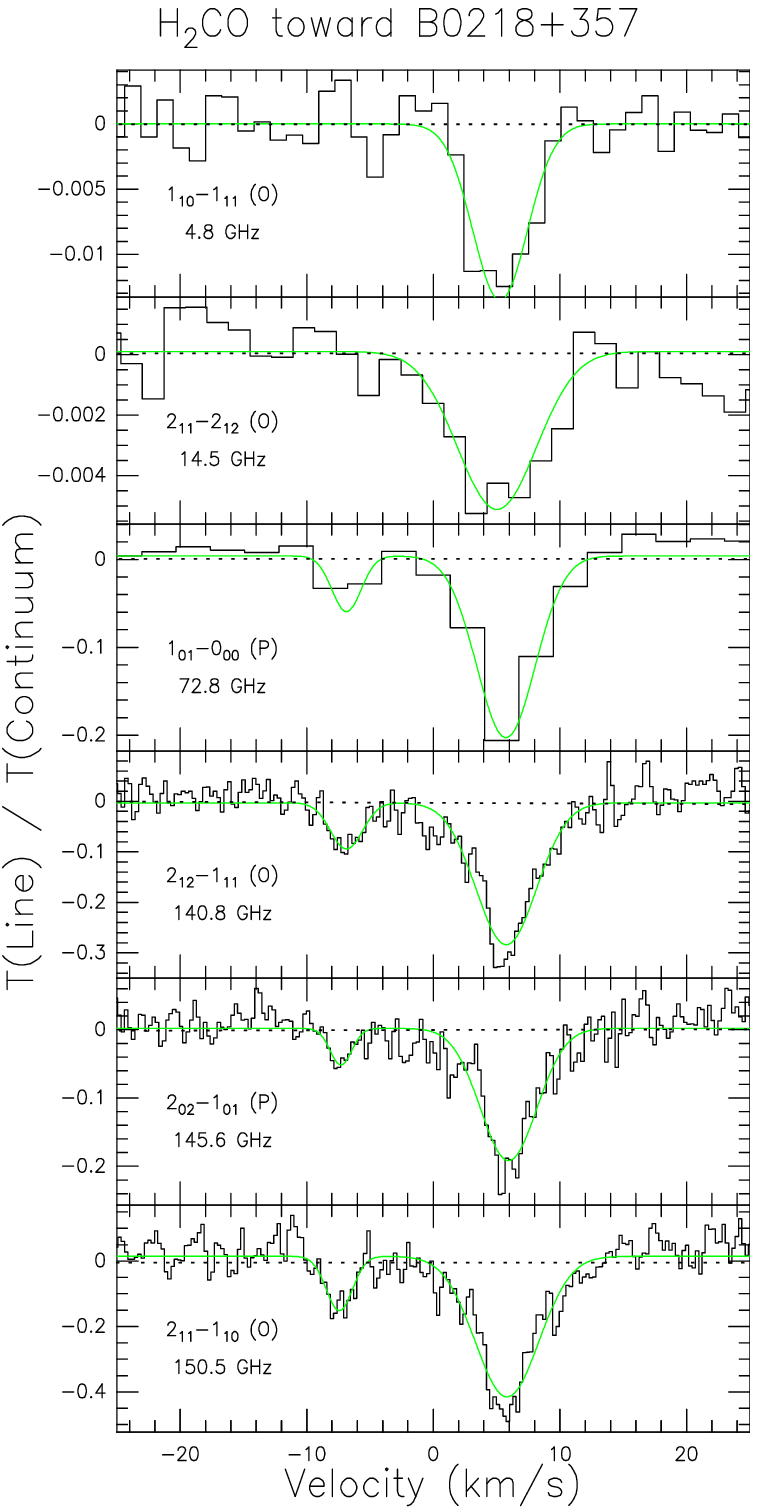}
\vspace{-0.0cm}
\caption{H$_2$CO lines observed toward B0218+357 and Gaussian fits. Given are line-to-continuum ratios (the entire continuum flux of B0218+357 
was taken) as a function of heliocentric velocity adopting a redshift of $z$=0.68466. The labels in the lower left corner of each spectrum show 
transition and rest frequency. O and P refer to ortho- and para-H$_2$CO, respectively.  
\label{7353fig3}}
\end{figure}

\subsection{Formaldehyde absorption toward image A}

All five formaldehyde lines we searched for are detected. Including the re-reduced $2_{11}-2_{12}$ line reported by Menten \& Reid (\cite{menten96}),
Fig.\,\ref{7353fig3} shows all H$_2$CO lines measured to date toward B0218+357. The upper two panels show "K-doublet" lines. Rotational lines are shown
in the lower four panels. In order to minimize systematic calibration errors and to focus on physical parameters, line-to-continuum ratios are given 
that provide a direct estimate of optical depths. Such ratios should also be free of problems related to variability of the calibrator, to variability 
of B0218+357 itself (e.g., Biggs et al. \cite{biggs01a, biggs01b}) or to elevation dependent gain variations of the respective telescope(s). 

Two of the measured H$_2$CO transitions belong to the para and four to the ortho species (Fig.\,\ref{7353fig3}). Systematic differences between para 
and ortho-H$_2$CO lines are not apparent. Instead, there are obvious differences between the two K-doublet profiles measured at cm-wavelengths 
and the four rotational transitions observed at mm-wavelengths. The former show line-to-continuum ratios that are far below unity, while the latter are 
characterized by much higher line-to-continuum ratios. And while the former only show one velocity component, centered at about +6\,km\,s$^{-1}$, 
the latter show a second component at about --7\,km\,s$^{-1}$. It is not clear whether this difference is intrinsic to the source or whether 
it is merely an effect of limited signal-to-noise ratios in the K-doublet lines. Line parameters obtained by Gaussian fits are given in Table~\ref{lines}.

Menten \& Reid (\cite{menten96}) found that absorption in the 2$_{11}-2_{12}$ line of ortho-H$_2$CO is confined to source A. The 1$_{01}-0_{00}$ 
para-H$_2$CO transition, that was taken with even higher angular resolution (see Table~\ref{obs}), yields the same result. There is no hint of 
line absorption toward source B with a 3.1\,km\,s$^{-1}$ channel 3$\sigma$ upper limit to the line-to-continuum ratio of 0.39 (see also
Muller et al. \cite{muller07}).

\begin{table*}[t]
\caption{H$_2$CO line parameters$^{a)}$}
\label{lines}
\begin{center}
\begin{tabular}{ccrcrcccc}
\hline
Line& Species&\multicolumn{1}{c}{Rest}&\multicolumn{1}{c}{$\int$$(T_{\rm L}$/$T_{\rm c}$)d$V$}&\multicolumn{1}{c}{$V_{\rm hel}$}& $\Delta V_{1/2}$ &
$|T_{\rm L}|$/$T_{\rm c}$ & $S_{\rm A}$/$S_{\rm tot}$ & Optical  \\
    &        &\multicolumn{1}{c}{frequency}&   & & & & & depth                                                                                    \\
    &        &  (GHz)                      &\multicolumn{3}{c}{(km\,s$^{-1}$)}  \\
\hline   \\

1$_{10}$$-$1$_{11}$ &O&   4.82 & --0.073$\pm$0.006 &   5.3$\pm$0.2 & 5.1$\pm$0.5 & 0.014$\pm$0.002 & 0.56 &  0.025$\pm$0.006 \\
2$_{11}$$-$2$_{12}$ &O&  14.48 & --0.038$\pm$0.004 &   5.1$\pm$0.4 & 8.7$\pm$0.9 & 0.004$\pm$0.0006& 0.70 &  0.006$\pm$0.002 \\   

1$_{01}$$-$0$_{00}$ &P&  72.83 & --1.154$\pm$0.097 &   5.7$\pm$0.2 & 5.3$\pm$0.5 & 0.217$\pm$0.028 & 0.77 &  0.331$\pm$0.047  \\ 
                    & &        & --0.175$\pm$0.066 & --6.9$\pm$0.4 & 2.7$\pm$0.5 & 0.065$\pm$0.027 &      &  0.088$\pm$0.036  \\

2$_{12}$$-$1$_{11}$ &O& 140.83 & --1.721$\pm$0.063 &   5.8$\pm$0.1 & 5.7$\pm$0.3 & 0.300$\pm$0.018 & 0.77 &  0.493$\pm$0.051   \\
                    & &        & --0.304$\pm$0.046 & --6.9$\pm$0.2 & 3.1$\pm$0.6 & 0.098$\pm$0.024 &      &  0.136$\pm$0.035  \\

2$_{02}$$-$1$_{01}$ &P& 145.60 & --1.047$\pm$0.052 &   6.0$\pm$0.1 & 5.1$\pm$0.4 & 0.206$\pm$0.018 & 0.77 &  0.311$\pm$0.034  \\
                    & &        & --0.105$\pm$0.029 & --7.3$\pm$0.3 & 1.9$\pm$0.7 & 0.056$\pm$0.026 &      &  0.076$\pm$0.034  \\  

2$_{11}$$-$1$_{10}$ &O& 150.49 & --2.458$\pm$0.102 &   5.8$\pm$0.1 & 5.4$\pm$0.3 & 0.454$\pm$0.028 & 0.77 &  0.890$\pm$0.071  \\
                    & &        & --0.363$\pm$0.056 & --7.4$\pm$0.2 & 2.2$\pm$0.4 & 0.164$\pm$0.037 &      &  0.240$\pm$0.053  \\
\hline   
\end{tabular}
\end{center}
a) Cols.\,1 and 2: H$_2$CO transition (O: ortho; P: para). Col.\,3: Rest frequency. Cols.\,4--7: Integrated line-to-continuum ratios, heliocentric
velocity with respect to $z$=0.68466, linewidth, and peak line-to-continuum ratio. The latter is calculated dividing the velocity integrated 
line-to-continuum ratios by the linewidth. The errors given are formal standard deviations that do not include calibration uncertainties. Col.\,8: 
Continuum flux density of image A w.r.t. the total continuum flux density. The error is about $\pm$0.03. Col.\,9: Peak "apparent" 
optical depth, i.e. $\tau$ = --ln(1--$|T_{\rm L}|$/$T_{\rm c}$), assuming uniform coverage of image A and an H$_2$CO excitation temperature 
that is small w.r.t. the continuum background. 
\end{table*}

\section{Discussion}

\subsection{Radio continuum maps and spectral indices}

Table~\ref{cont} indicates that the flux density of B0218+357 reaches a peak near 10\,GHz. However, B0218+357 is a variable source (e.g., 
Biggs et al. \cite{biggs01b}) and our data were taken during a time interval of two years. In order to obtain a reliable radio continuum 
spectrum, quasi-simultaneous observations at a number of frequencies are mandatory. Mittal et al. (\cite{mittal06}) presented such observations 
at five frequencies ranging from 1.65 to 15.35\,GHz. The combined flux density of sources A and B (excluding the Einstein ring) reaches indeed 
a peak near 10\,GHz (0.8--0.9\,Jy; Mittal et al. \cite{mittal07}). Our total fluxes (Table~\ref{cont}) include the ring and are thus higher. 

The frequency range covered here by the continuum maps is larger than that of Mittal et al. (\cite{mittal06}). Expecting systematically decreasing 
flux densities beyond $\sim$15\,GHz, our $\sim$85\,GHz fluxes are consistent with the corresponding part of the continuum spectrum shown by Combes 
\& Wiklind (\cite{combes97a}) and Combes et al. (\cite{combes97b}) that suggest a spectral index of $\alpha$ $\sim$ --0.25 ($S_{\nu}$ $\propto$ 
$\nu^{\alpha}$). The minimum in our "spectrum" at 43.2\,GHz (Table~\ref{cont}) may be caused by time variability or by a substantial amount 
of missing flux. Among the data presented here, the 43.2\,GHz Q-band map (Fig.\,2) has the highest angular resolution so that all spatial 
components except images A and B are resolved out. 

This might also affect our measured A/B flux density ratios. While B is located near the center of the Einstein ring, A is placed near its 
northeastern edge. Thus estimated flux densities for B may be contaminated by Einstein ring emission at 8.6 and 14.1\,GHz, but not at 43.2\,GHz. 
This would be consistent with lower ratios at the two lower frequencies. Nevertheless, our A/B flux density ratios, $\sim$ 3.3 (at 8.6 and 14.1\,GHz) 
and $\sim$3.5 (at 43.2\,GHz) are, within the formal errors (see Table~\ref{cont}), consistent with a single value. Furthermore, they also lie 
within the range of values obtained by Mittal et al. (\cite{mittal06}; their Table~3 and Fig.\,5b) for frequencies 8.4--15.4\,GHz. Thus we confirm 
that there is no evidence for a significant amount of free-free absorption between 8.6 and 43.2\,GHz (see Mittal et al. \cite{mittal07} for a model 
of such an ionized absorbing layer).

\subsection{Spectral analysis}

\subsubsection{General aspects}

Table~\ref{lines} gives integrated (Col.\,4) and peak (Col.\,7) line-to-continuum ratios of the six lines shown in Fig.\,\ref{7353fig3} as well 
as the fractions of the total continuum flux originating from image A (Col.\,8) and the resulting peak apparent optical depths (Col.\,9). Assuming 
that only image A is absorbed by a foreground molecular cloud (see Sects.\,1 and 3.1 and Mittal et al. \cite{mittal07}) and that the coverage is 
uniform, the rotational H$_2$CO transitions have optical depths 0.3 $<$ $\tau$ $<$ 0.9, while the K-doublet opacities are at least an order of 
magnitude lower. 

Henkel et al. (\cite{henkel05}) interpreted the trend of increasing optical depth of molecular lines with rising frequency in terms of a uniform 
molecular absorber located toward the line-of-sight to the peak of a radio continuum source. In this scenario, the size of the continuum source 
decreases with increasing frequency. In view of the trend in apparent optical depth that is obvious when comparing lines at \hbox{dm-}, cm- and 
mm-wavelengths (e.g., Carilli et al. \cite{carilli93}; Menten \& Reid \cite{menten96}; Wiklind \& Combes \cite{wiklind95}) it may seem to be 
surprising that the $J$=2--1 para-H$_2$CO line, observed at a similar frequency as the corresponding ortho-lines is less optically thick 
(Table~\ref{lines}). The difference may be caused by a high ortho- to para-H$_2$CO abundance ratio or by a low excitation of the absorbing 
molecular gas (see Combes et al. \cite{combes97b} for CS excitation and Combes \& Wiklind \cite{combes97a} and Gerin et al. \cite{gerin97} for 
limits to the H$_2$O and $^{13}$CO excitation). For ortho-H$_2$CO, the $J$=2--1 lines are the lowest rotational transitions. Para-H$_2$CO, however, 
also has a $J$=1--0 line. Thus the  $J$=2--1 para-H$_2$CO line is not the lowest rotational line in its respective $K_{\rm a}$ ladder 
(Fig.\,\ref{7353fig1}) and Table~\ref{lines} shows indeed that the $J$=1--0 line has an optical depth that is similar to that of the $J$=2--1 
line, in spite of the lower statistical weights of its two rotational states. However, it has to be emphasized that the assumption of a uniform 
molecular cloud is highly unrealistic. The use of this assumption is only justified by the limited spatial resolution available when studying the 
molecular content of a galaxy at redshift $z$=0.68.

\subsubsection{Continuum source morphology}

Sensitive high resolution continuum maps of image A at 8.4\,GHz show a source of size $\sim$10$\times$10\,mas$^2$ that is edge-brightened on its 
south-western side (Fig.\,2 of Biggs et al. \cite{biggs03}). Tangential stretching by a factor of 3--4 along position angle P.A. $\sim$ --30$^{\circ}$ 
almost matches the elongation caused by a jet at P.A. $\sim$ +60$^{\circ}$. At low frequencies ($<$8\,GHz), the morphology of component A may be 
affected by free-free absorption (Mittal et al. \cite{mittal07}). At higher frequencies, VLBI continuum measurements of B0218+357 have been made up 
to 22 and 43\,GHz (Porcas \& Patnaik \cite{porcas96}; Porcas \cite{porcas04}), but sensitivities are not high enough to reveal additional spatial fine 
structure. It would not be a surprise if the jet contribution would fade at higher frequencies leaving the tangentially stretched core that may coincide 
in position with a spiral arm of the lensing galaxy (see Fig.\,3 of York et al. \cite{york05}). Assuming that the core shows a brightness temperature 
and flux density not drastically varying with frequency in the 10--100\,GHz frequency range (e.g., Blandford \& K{\"o}nigl \cite{blandford79}), its 
solid angle would vary as $\Omega$ $\propto$ $\nu^{-2...-1}$ (see also Lobanov \cite{lobanov98}) between cm and mm-wavelengths. With a spectral 
index of $\alpha$ $\sim$ --0.25 (Sect.\,4.1) the mm-wave flux of B0218+357 is about half of that at cm-wavelengths, which is compatible with a fading 
jet and a flat spectrum core.  

\subsubsection{LVG simulations}

To test the qualitative scenario outlined in Sect.\,4.2.1 and to account for the background continuum morphology sketched in Sect.\,4.2.2, we use Large 
Velocity Gradient (LVG) models for para- and ortho-H$_2$CO, accounting for the 41 and 40 lowest energy levels of each species up to 300\,cm$^{-1}$ 
above the ground state. We adopt collision rates from Green (\cite{green91}) and assume a spherical cloud geometry. The choice of a particular cloud 
geometry can affect the resulting densities up to half an order of magnitude, but only if the lines are optically thick. Applying a plane-parallel 
instead of a spherical cloud geometry, resulting particle densities could be lower by up to this amount. 

Since line shapes are identical within the limits of noise, we can try to reproduce the various peak optical depths of the main velocity component 
displayed in Table~\ref{lines}. As we shell see, main constraints are obtained with the four ortho-H$_2$CO transitions, while the para-H$_2$CO lines serve as 
a measure of ortho-to-para abundance ratios. 

\hfill\break\noindent
The simplest case, ignoring collisions: \\
As a first approach, we neglect collisional excitation as well as frequency dependent variations of the underlying continuum morphology (Sect.\,4.2.2). 
Since the data do not significantly constrain the temperature of the cosmic microwave background (CMB), we adopt $T_{\rm CMB}$ = 2.73\,$\times$\,(1+$z$)\,K 
= 4.60\,K = $T_{\rm ex}$ ($T_{\rm ex}$: excitation temperature). Table~\ref{tau} displays the resulting optical depths (Model~I) with minimized reduced 
$\chi^2$ values of 7.1 for the four ortho- and 1.7 for the two para-H$_2$CO transitions ($\chi^2$/($N-P$), $N$: number of lines, $P$: number 
of free parameters; $P_{\rm ortho}$ = 2 (density and column density), $P_{\rm para}$ = 1 (column density)). The total column density is $N$(H$_2$CO) $\sim$ 
3.9$\times$10$^{13}$\,cm$^{-2}$ and the ortho to para abundance ratio becomes $\sim$2.2. Modeled optical depths of the two para-H$_2$CO transitions 
are well within the 2$\sigma$ limits of the observational errors. For the two cm-wave K-doublet ortho-lines (see Fig.\,\ref{7353fig1}), however, modeled 
opacities appear to be too high, while for the mm-wave $J$=2--1 lines modeled optical depths tend to be too low. Therefore the calculated ortho-values raise 
some doubts about the quality of the fit.

\begin{table}[t]
\caption{Optical depths$^{\rm a)}$}
\label{tau}
\begin{center}
\begin{tabular}{lccc}
\hline
Line & Observed & Model & Model \\
     &          &   I   &  II   \\ 
\hline   \\
1$_{10}$$-$1$_{11}$~(O)          & 0.025$\pm$0.007 & 0.028 & 0.036 \\
2$_{11}$$-$2$_{12}$~(O)          & 0.006$\pm$0.002 & 0.010 & 0.007 \\   
2$_{12}$$-$1$_{11}$~(O)          & 0.493$\pm$0.090 & 0.436 & 0.183 \\
2$_{11}$$-$1$_{10}$~(O)          & 0.890$\pm$0.151 & 0.426 & 0.160 \\
\hline 
1$_{01}$$-$0$_{00}$~(P)          & 0.331$\pm$0.068 & 0.257 & 0.250 \\ 
2$_{02}$$-$1$_{01}$~(P)          & 0.311$\pm$0.057 & 0.352 & 0.346 \\
\hline \hline
$\chi^2_{\rm reduced, ortho}$    &     --          & 7.1   &18.9   \\
$\chi^2_{\rm reduced, para}$     &      --         & 1.7   & 1.8   \\
$N$(H$_2$CO)                     &      --         & 3.9   & 2.6   \\
Ortho/Para                       &      --         & 2.2   & 0.9   \\
\hline

\end{tabular}
\end{center}
a) Col.\,1: H$_2$CO transition with O and P denoting ortho- and para-H$_2$CO, respectively. The upper two lines refer to K-doublet transitions. In the lower
part of the Table, reduced $\chi^2$ values, total H$_2$CO column densities from the +7\,km\,s$^{-1}$ component in units of 10$^{13}$\,cm$^{-2}$, and H$_2$CO 
ortho-to-para abundance ratios are given. \\
Col.\,2: Observed apparent optical depths taken from the last column of Table~\ref{lines}. Given errors are larger than those in Table~\ref{lines}, also 
accounting for a calibration uncertainty of $\pm$15\%. \\
Col.\,3: LVG results exclusively accounting for microwave background excitation ($T_{\rm CMB}$ = 4.6\,K = $T_{\rm ex}$; Model~I). \\
Col.\,4: Same as Col.\,3, but including collisional excitation ($n$(H$_2$) = 500\,cm$^{-3}$; Model~II). For details, see Sect.\,4.2.3. \\
\end{table}

\begin{table}[t]
\caption{Optical depths accounting for source coverage$^{\rm a)}$}
\label{taufc}
\begin{center}
\begin{tabular}{lcccc}
\hline
Line & Observed & Model & Model & Model \\
     &          &  III  &   IV  &   V   \\ 
\hline   \\
1$_{10}$$-$1$_{11}$~(O)         & 0.113$\pm$0.032 & 0.046 & 0.086 & 0.120 \\
2$_{11}$$-$2$_{12}$~(O)         & 0.016$\pm$0.006 & 0.017 & 0.020 & 0.024 \\   
2$_{12}$$-$1$_{11}$~(O)         & 0.539$\pm$0.098 & 0.726 & 0.641 & 0.610 \\
2$_{11}$$-$1$_{10}$~(O)         & 0.941$\pm$0.160 & 0.710 & 0.591 & 0.533 \\
\hline
1$_{01}$$-$0$_{00}$~(P)         & 0.503$\pm$0.103 & 0.286 & 0.285 & 0.283 \\ 
2$_{02}$$-$1$_{01}$~(P)         & 0.335$\pm$0.061 & 0.393 & 0.393 & 0.392 \\
\hline \hline  
$\chi^2_{\rm reduced, ortho}$   &        --       & 4.9   & 3.5   & 4.4   \\ 
$\chi^2_{\rm reduced, para}$    &        --       & 5.3   & 5.4   & 5.4   \\ 
$N$(H$_2$CO)                    &        --       & 5.4   & 5.2   & 4.8   \\
Ortho/Para                      &        --       & 3.0   & 2.8   & 2.5   \\
\hline
\end{tabular}
\end{center}
a) Col.\,1: H$_2$CO transition with O and P denoting ortho- and para-H$_2$CO, respectively. The upper two lines refer to K-doublet transitions. In the lower
part of the Table, reduced $\chi^2$ values, total H$_2$CO column densities from the +7\,km\,s$^{-1}$ component in units of 10$^{13}$\,cm$^{-2}$, and H$_2$CO 
ortho-to-para abundance ratios are given. \\
Col.\,2: Observed apparent optical depths (see Table~\ref{tau}), but accounting for a frequency dependence of the continuum source covering factor, 
$f_{\rm c}$ = ($\nu_{\rm GHz}/100)^{0.5}$. \\
Col.\,3: LVG results exclusively accounting for microwave background excitation ($T_{\rm CMB}$ = 4.6\,K = $T_{\rm ex}$; Model~III). \\
Cols.\,4 and 5: Same as Col.\,3, but including collisional excitation with $n$(H$_2$) = 200 and 500\,cm$^{-3}$, respectively (Models~IV and V). For details, 
see Sect.\,4.2.3.
\end{table}

\hfill\break\noindent
Introducing collisional processes: \\
Toward B0218+357, neglecting collisions is a potentially realistic assumption, when analyzing diatomic or linear molecules with a CO-like energy level 
scheme (see Combes \& Wiklind \cite{combes97a}, Gerin et al. \cite{gerin97}, and Henkel et al. \cite{henkel05}), H$_2$CO, however, critically responds to
collisional excitation even at very low densities. This is a consequence of the collisional cooling of the K-doublet lines (see e.g., Garrison et al. 
\cite{garrison75}) that sets in at densities as low as $n$(H$_2$)$\sim$100\,cm$^{-3}$, while collisional heating of the rotational transitions is still far 
from being significant. Collisional cooling of the 1$_{10}-1_{11}$ and 2$_{11}-2_{12}$ lines is rather independent of the chosen kinetic temperature as long 
as $T_{\rm kin}$ $>$ 10\,K. Here we adopt $T_{\rm kin}$ = 55\,K (Henkel et al. \cite{henkel05}). Reduced $T_{\rm ex}$ values cause increased optical 
depths in the K-doublet lines. As a consequence, the optical depths of the 1$_{10}-1_{11}$ K-doublet and the rotational ortho-H$_2$CO lines become even more 
similar than in model~I and a reasonable fit cannot be obtained. This is exemplified for a density of $n$(H$_2$) = 500\,cm$^{-3}$ in Model~II of Table~\ref{tau},
which is characterized by a reduced $\chi^2$ value of 18.9 for the ortho-transitions. Higher densities, i.e. 500\,cm$^{-3}$ $<$ $n$(H$_2$) $<$ 10$^5$\,cm$^{-3}$, 
pose similar problems, while densities $\ga$10$^5$\,cm$^{-3}$ severely contradict additional constraints that have been discussed by Combes et al. 
(\cite{combes97a}), Gerin et al. (\cite{gerin97}), and Henkel et al. (\cite{henkel05}). 

\hfill\break\noindent
A frequency dependent radio continuum morphology: \\
In view of the problems outlined above, the best way to reconcile observational and model data is to account for the frequency dependence of the continuum 
morphology (Sect.\,4.2.2). A background continuum source covering factor $f_{\rm c}$, increasing with frequency, would (1) reduce the differences in optical depths 
between the 1$_{10}-1_{11}$ K-doublet and $J$=2--1 rotational lines and would (2) raise the optical depth ratio of the K-doublet lines. At 137\,GHz, $f_{\rm c}$$\sim$1 
for image A (Wiklind \& Combes \cite{wiklind95}). At 0.85\,GHz, we get from the redshifted 1.7\,GHz OH absorption lines and the continuum flux of the entire source 
(Kanekar et al. \cite{kanekar03}) a peak apparent optical depth of 0.01, implying that the covering factor $f_{\rm c}$ is $\ga$0.01. Assuming that $f_{\rm c}$ 
= 1 is reached at exactly 100\,GHz, this corresponds to $f_{\rm c}$ $\propto$ $\nu^{0.97}$ (choosing instead 70 or 130\,GHz would yield similar exponents, i.e. 1.04 
or 0.91). The lower $f_{\rm c}$ limit at 0.85\,GHz is, however, too small. Firstly, the 1.667 and 1.665\,GHz OH lines show the intensity ratio expected in the 
optically thin limit at local thermodynamic equilibrium conditions (1.8:1). Therefore, unlike for CO, the OH optical depths are well below unity ($\tau$$\la$0.5)}. 
This increases the attainable line-to-continuum ratio in the optically thick case (i.e. the source covering factor) at 0.85\,GHz by at least a factor of two and 
yields $f_{\rm c}$ $\propto$ $\nu^{0.0 ... 0.8}$. Secondly, we also have to account for the fact that at low frequencies the Einstein ring becomes 
prominent and that image A is not dominating the total flux density (for free-free absorption of image A, see Mittal et al. \cite{mittal06}). Only taking the 
continuum flux from image A thus further raises the 0.85\,GHz line to continuum ratio and $f_{\rm c}$ $\propto$ $\nu^{0.0 ... 0.5}$ becomes a more realistic 
estimate. This is a weaker frequency dependence than the hypothesized one for $\Omega$, the continuum solid angle (Sect.\,4.2.2), but agrees with the observed 
overall spectral index of the continuum radiation. 

In order to study the case with an exponent for $f_{\rm c}$ of 0.5 that differs most from a scenario with a frequency independent $f_{\rm c}$ value, 
Table~\ref{taufc} displays the $f_{\rm c}$-corrected observed optical depths as well as LVG model results for densities $n$(H$_2$) = 0, 200, and 500\,cm$^{-3}$. 
Excluding collisions (Model~III), the two K-doublet lines show optical depths that are too similar for a good fit. Minimized reduced $\chi^2$ values are 
4.9 and 5.3 for the ortho and para-H$_2$CO lines, respectively. A density of 200\,cm$^{-3}$ (Model~IV) reproduces much better the K-doublet lines and 
shows with reduced $\chi^2$ values of 3.5 and 5.4 the highest probablity. Total H$_2$CO column densities and ortho-to-para abundance ratios become 
5.2$\times$10$^{13}$\,cm$^{-2}$ and 2.8. At higher densities we get the same effect as in the case of a frequency independent $f_{\rm c}$ value. Opacities of K-doublet and 
rotational ortho-H$_2$CO transitions slowly converge and $\chi^2$ values gradually increase with rising density. Model~V shows this for $n$(H$_2$) = 500\,cm$^{-3}$, 
while for $n$(H$_2$) = 1000 and 3000\,cm$^{-3}$, reduced $\chi^2$ values for the ortho-H$_2$CO lines become 6.0 and 8.8, respectively. The significant difference 
in the measured optical depths of the two $J$=2--1 lines of ortho-H$_2$CO is not reproduced in any model and deserves observational confirmation. 

From Table~\ref{taufc} it is apparent that the fits to the para-H$_2$CO lines are almost independent of density, because rotational excitation is minimal at 
the densities considered here. The LVG simulations of the para-H$_2$CO lines in models III, IV, and V are not as good as in models I and II. While obviously not 
being suitable for a reliable $T_{\rm CMB}$ determination, calculated opacities are still within the 1 and 3$\sigma$ limits of the observed values for the 
2$_{02}-1_{01}$ and 1$_{01}-0_{00}$ lines, respectively.

\subsubsection{More general implications}

To summarize, the successful models (I, III, V, and particularly IV) agree well with each other, both with respect to column density and ortho-to-para abundance 
ratio. The density of the gas is $n$(H$_2$) $<$ 10$^{3}$\,cm$^{-3}$, most likely $\sim$200\,cm$^{-3}$, supporting the view that the cloud is diffuse and has 
peculiar physical parameters rarely found in the nearby universe (cf., Henkel et al. \cite{henkel05}). For the best scenario, $T_{\rm kin}$ = 55\,K (from NH$_3$) 
and $n$(H$_2$) = 200\,cm$^{-3}$ (from H$_2$CO), we have calculated excitation temperatures, assuming a cosmic background radiation of 4.6\,K. In the optically thin case, 
excitation temperatures of the three ground rotational transitions of CS, HCN, HNC, HCO$^+$, and N$_2$H$^+$ are within 1\% of the assumed CMB background temperature 
and should thus be extremely useful for future measurements aiming at $T_{\rm CMB}$ as a function of redshift. Rare CO isotopomers, however, are not as useful.

There is a second velocity component, at \hbox{--7}\,km\,s$^{-1}$ (see Fig.\,\ref{lines} and Table~\ref{lines}), that is also apparent (mostly in the form of 
skewed profiles) in transitions of other molecules (Wiklind \& Combes \cite{wiklind95}; Combes et al. \cite{combes97b}; Combes \& Wiklind \cite{combes98}; Kanekar 
et al. \cite{kanekar03}). The broad component seen in ammonia (NH$_3$) at a heliocentric velocity of $\sim$0\,km\,s$^{-1}$ (Henkel et al. \cite{henkel05} use Local 
Standard of Rest (LSR) velocities, while here heliocentric velocities are given) is a superposition of hyperfine satellites of the $\sim$ +6\,km\,s$^{-1}$ and the 
entire \hbox{--7}\,km\,s$^{-1}$ line and may thus serve as a rough measure of the relative importance of the two kinematic NH$_3$ features. As is the case for 
ammonia, the \hbox{--7}\,km\,s$^{-1}$ component of formaldehyde is too weak for detailed modeling but adds $\sim$15\% to the resulting column densities given 
in Tables~\ref{tau} and \ref{taufc}. 

The ortho-H$_2$CO abundance appears to be slightly larger than that estimated by Menten \& Reid (\cite{menten96}). Having carried out multiline studies
of ammonia (NH$_3$) and formaldehyde toward B0218+357, we can now compare their abundances. For the total column densities (because of hyperfine splitting, 
the components are not well separated in NH$_3$) we find $N$(H$_2$CO) $\sim$ 0.6 $\times$ $N$(NH$_3$) (for NH$_3$, Henkel et al. \cite{henkel05}). The 
$N$(H$_2$CO)/$N$(NH$_3$) abundance ratio is four times smaller than the average ratio estimated for local diffuse clouds observed in absorption 
(Liszt et al. \cite{liszt06}). In view of a possible overabundance of the more `primary' $^{12}$C and $^{16}$O nuclei relative to the mainly `secondary' 
$^{14}$N nucleus in spiral galaxies of the distant past (see e.g., Wheeler et al. \cite{wheeler89} for a review on metallicity dependent CNO abundances) 
this may seem to be surprising. However, the galactic H$_2$CO abundances are based on two H$_2$CO transitions (versus six here), while NH$_3$ data are from 
measurements of two spectral lines only (versus four in the case of B0218+357). For only a few diffuse clouds galactic NH$_3$ absorption profiles have been 
measured. Therefore the `local' average ratio (see also Nash \cite{nash90}) may be less certain than the ratio presented here for a single distant gravitational 
lens alone. While our H$_2$CO column density is 2.5 times higher than the highest column density among the sources summarized by Liszt et al. (\cite{liszt06}), 
our absorption profile is wider by at least this amount, thus yielding a similar column density per km\,s$^{-1}$ velocity interval. If H$_2$CO is preferentially 
formed on dust grains while NH$_3$ is arising in the gas phase from either N$^+$ or N (e.g., Liszt et al. \cite{liszt06}), a good correlation between the 
abundances of the two molecules may not be expected. 

An ortho-to-para abundance ratio of almost three and a kinetic temperature of 55\,K (Henkel et al. \cite{henkel05}) are consistent with the correlation 
between kinetic temperature and ortho-to-para abundance ratio shown in Fig.\,10 of Kahane et al. (\cite{kahane84}). Whether this indicates gas phase H$_2$CO 
formation in a warm medium or whether adsorption and desorption on grain surfaces can lead to efficient thermalization (Mangum \& Wootten \cite{mangum93}) 
remains to be seen.

\section{Conclusions}

The detection of six H$_2$CO absorption lines toward the gravitational lens B0218+357 reveals surprisingly detailed information on molecular gas at redshift
$z$=0.68. Our analysis leads to the following main results:

\begin{itemize}

\item The continuum flux density ratio of component A to B is $\sim$3.3 between 8.6 and 43\,GHz, does not vary significantly, and is thus not severely
affected by selective free-free absorption in this frequency range.

\item The 1$_{01}-0_{00}$ line of para-H$_2$CO is mainly absorbing the radio continuum from image A, consistent with previously published results from
the 2$_{11}-2_{12}$ line of ortho-H$_2$CO.

\item The H$_2$CO column density of the main spectral feature is $\sim$5$\times$10$^{13}$\,cm$^{-2}$. The mm-wave lines show a second, blue-shifted 
velocity component that is displaced by about 13\,km\,s$^{-1}$ from the main component. It contributes another $\sim$15\% to the total column density.

\item The best fits to the ortho-H$_2$CO molecular lines involve the assumption of a continuum cloud coverage that, between 0.85 and 100\,GHz, is 
likely frequency dependent ($f_{\rm c}$ $\propto$ $\nu^{0.0...0.5}$). 

\item The gas density is $n$(H$_2$) $<$ 1000\,cm$^{-3}$, most likely $\sim$200\,cm$^{-3}$, confirming the view that the absorption is arising from 
a diffuse molecular environment.

\item With $T_{\rm kin}$ and $n$(H$_2$) known, optically thin lines of the three ground rotational transitions of CS, HCN, HNC, HCO$^+$, and N$_2$H$^+$
(not CO) will be ideal measures of the cosmic microwave background at $z$=0.68.

\item The total H$_2$CO column density is about 60\% of that obtained for NH$_3$. This is four times below the average ratio deduced from local 
galactic diffuse gas seen in absorption. 

\item The ortho-to-para H$_2$CO abundance ratio is large (2.0--3.0), reflecting the kinetic temperature of the cloud. Whether this is caused by cloud 
formation in a warm medium or whether this is due to efficient thermalization on dust grain mantles remains open.

\end{itemize}

\begin{acknowledgements}

We wish to thank C.M. Walmsley and an unknown referee for useful comments. This research has made use of NASA's Astrophysical Data System. N. Jethava was supported for this research through a stipend from the International Max Planck Research School (IMPRS) for Radio and Infrared Astronomy at the Universities of Bonn and Cologne. 

\end{acknowledgements}

\end{document}